\documentclass[10pt,conference]{IEEEtran}

\usepackage{cite}
\usepackage{amsmath,amssymb}
\usepackage{array}
\usepackage{booktabs}
\usepackage{enumitem}
\usepackage{graphicx}
\usepackage{longtable}
\usepackage{textcomp}
\usepackage{multirow}
\usepackage[T1]{fontenc}
\let\paperdefaulttt\ttdefault
\usepackage[scaled=0.92]{newtxtt}
\renewcommand{\ttdefault}{\paperdefaulttt}
\usepackage{url}
\usepackage[table]{xcolor}
\usepackage{tikz}
\usetikzlibrary{arrows.meta,fit,positioning,calc,shapes.geometric}
\usepackage{xstring}
\usepackage{upquote}
\usepackage{listings}
\usepackage[most]{tcolorbox}
\usepackage[hidelinks]{hyperref}

\newif\ifshowcomments
\showcommentstrue
\ifshowcomments
  \newcommand{\jie}[1]{\textcolor{blue}{\textbf{[Jie: #1]}}}
  \newcommand{\gunel}[1]{\textcolor{red}{\textbf{[Gunel: #1]}}}
  \newcommand{\czp}[1]{\textcolor{orange}{\textbf{[ZP: #1]}}}
\else
  \newcommand{\jie}[1]{}
  \newcommand{\gunel}[1]{}
  \newcommand{\czp}[1]{}
\fi
\newtcolorbox{rqanswer}[1]{%
  enhanced,
  colback=gray!20,
  colframe=black,
  boxrule=1pt,
  arc=1.5pt,
  left=6pt,right=6pt,top=4pt,bottom=4pt,
  before upper={\textbf{#1:}\space},
}

\DeclareRobustCommand{\breaktt}[1]{%
  \begingroup
  \ttfamily
  \hyphenpenalty=3000%
  \exhyphenpenalty=3000%
  \StrSubstitute{#1}{-}{\discretionary{-}{}{-}}%
  \endgroup
}

\lstdefinestyle{worked}{%
  basicstyle=\footnotesize\ttfamily,
  breaklines=true,
  breakatwhitespace=true,
  breakindent=2em,
  columns=fullflexible,
  keepspaces=true,
  showstringspaces=false,
  xleftmargin=0pt,
  xrightmargin=0pt,
  frame=none,
  aboveskip=2pt,
  belowskip=2pt,
  escapeinside={(*@}{@*)},
}
\lstdefinestyle{prompt}{%
  basicstyle=\scriptsize\ttfamily,
  breaklines=true,
  breakatwhitespace=true,
  breakautoindent=false,
  breakindent=0pt,
  columns=fullflexible,
  keepspaces=true,
  showstringspaces=false,
  xleftmargin=0pt,
  xrightmargin=0pt,
  frame=none,
  aboveskip=0pt,
  belowskip=0pt,
  escapeinside={(*@}{@*)},
}

\def\BibTeX{{\rm B\kern-.05em{\sc i\kern-.025em b}\kern-.08em
    T\kern-.1667em\lower.7ex\hbox{E}\kern-.125emX}}

\makeatletter
\long\def\@makecaption#1#2{%
\ifx\@captype\@IEEEtablestring%
\footnotesize\bgroup\par\centering\@IEEEtabletopskipstrut{\normalfont\footnotesize {#1.}\nobreakspace #2}\par\addvspace{0.5\baselineskip}\egroup%
\@IEEEtablecaptionsepspace
\else
\@IEEEfigurecaptionsepspace
\setbox\@tempboxa\hbox{\normalfont\footnotesize {#1.}\nobreakspace #2}%
\ifdim \wd\@tempboxa >\hsize%
\setbox\@tempboxa\hbox{\normalfont\footnotesize {#1.}\nobreakspace}%
\parbox[t]{\hsize}{\normalfont\footnotesize \noindent\unhbox\@tempboxa#2}%
\else%
\hbox to\hsize{\normalfont\footnotesize\hfil\box\@tempboxa\hfil}%
\fi\fi}
\makeatother

\begin{document}

\title{An Empirical Study of Downstream Adaptation\\for Agent Skills
}

\author{%
\IEEEauthorblockN{Xinjian Wu\IEEEauthorrefmark{1}, Jingzhi Gong\IEEEauthorrefmark{1}, Gunel Jahangirova\IEEEauthorrefmark{1}, Zhenpeng Chen\IEEEauthorrefmark{2}, Jie M. Zhang\IEEEauthorrefmark{1}}
\IEEEauthorblockA{\IEEEauthorrefmark{1}\textit{King's College London}, London, United Kingdom\\
\IEEEauthorrefmark{2}\textit{Tsinghua University}, Beijing, China\\
\{xinjian.wu, jingzhi.gong, gunel.jahangirova, jie.zhang\}@kcl.ac.uk,\quad zpchen@tsinghua.edu.cn}
}

\pagestyle{plain}
\maketitle
\thispagestyle{plain}

\begin{abstract}

As Large Language Model (LLM) agents become integral to modern software systems, ``skills'' have emerged as a novel unit of software reuse, enabling developers to package workflows, decision procedures, and prompt-based policies.
While skills are intended for reuse, downstream developers frequently modify published skills
to fit local contexts, yet little is known about the nature of such
adaptations. 
This paper presents the first empirical study
of downstream skill adaptation in public forks, to 
understand how published skills are adapted, and to  provide
implications for researchers and engineers on improving skill design, evolution,
and orchestration. 
Specifically, we analyze
1,126 skill-adaptation instances from six widely adopted skill repositories
and develop a taxonomy comprising 46 adaptation patterns organized
into 13 families. 
Our key findings reveal a reuse paradox: 
although skills are intended to be easily imported and reused,
developers spend a lot of effort rewriting what the skills do, fixing skill discoverability, 
and translating them for different tools and languages, indicating a need for better abstractions, standardized interfaces, and automated support for skill adaptation.
Furthermore, adaptations are highly interdependent, with changes in one component often requiring coordinated updates elsewhere, motivating automated support for detecting inconsistent modifications.
We also find that nearly one-fifth of adaptations introduce
security-sensitive content within the same instruction text that governs
behavior.
\end{abstract}






\section{Introduction}\label{sec:introduction}

Large language model (LLM) agents are increasingly extended through reusable
capabilities rather than only through task-specific prompts or tool calls.
Skills play a role analogous to software libraries: they package reusable
capabilities so that complex behavior does not need to be rebuilt from scratch
for each task. Public skill marketplaces already host more than 42,000
skills~\cite{liuAgentSkillsWild2026}, making them a prominent mechanism
for procedural knowledge to enter an LLM agent's context. A skill can
provide the agent with a domain workflow, a project-specific policy, a
review checklist, or a companion script to run, turning reusable knowledge
into operational guidance for concrete tasks.

Existing research generally treats a published skill as a reusable artifact
and can be organized around four closely related aspects~\cite{zhouComprehensiveSurveyAgent2026a}.
Skill representation studies how procedural knowledge is encoded and packaged
into reusable skills~\cite{zhouComprehensiveSurveyAgent2026a}. Skill
acquisition examines how skills are created from human expertise, agent
experience, task requirements, or external
resources~\cite{xuAgentSkillsLarge2026}. Skill retrieval and selection
investigates how agents discover, route, and compose skills from large
repositories~\cite{jiangSoKAgenticSkills2026,zhengSkillRouterSkillRouting2026,
liOrganizingOrchestratingBenchmarking2026a}. Skill evolution concerns how
published skills are evaluated, updated, validated, and maintained over
time~\cite{dingAgentSkillEvaluation2026}. Related work also evaluates skill
effectiveness on downstream tasks~\cite{hanSWESkillsBenchAgentSkills2026,
liSkillsBenchBenchmarkingHow2026} and identifies security risks in shared
skill repositories~\cite{liuAgentSkillsWild2026,liNoAttackRequired2026}.

However, the literature primarily examines skills as artifacts designed and
maintained by their original creators, leaving their downstream adaptation
largely unexplored. In practice, skills often encode assumptions about project
structure, available tools, organizational policies, supporting resources, and
expected workflows. Developers reusing a published skill in a different
environment may therefore need to extend, repair, personalize, or reorganize
its components (i.e., the \texttt{SKILL.md} instruction file and bundled
resources such as scripts and reference documents) to make it locally useful.
Such adaptations reveal where published skills do not transfer directly, which
components are most sensitive to context, and what needs recur across
independent users. Understanding these recurring adaptation needs is important
for improving skill design, evolution, and orchestration, yet little is known
about how developers adapt published skills after adoption or which adaptation
patterns occur systematically across downstream repositories.

To fill this gap, we study skill adaptation using public GitHub fork histories
as empirical evidence. We define a skill-adaptation instance as a downstream
branch snapshot in which a developer has introduced unique adaptations to an
upstream skill package with patterns labeled. Using six highly forked and widely adopted skill
repositories, we construct a corpus of 1,126 skill-adaptation instances and
develop a taxonomy of 46 adaptation patterns organized into 13 families. On
this basis, we investigate five research questions:

\begin{itemize}[leftmargin=*]
\item \textbf{RQ1 (Adaptation Patterns and Prevalence):} What adaptation patterns
  and families occur in public forks, and how prevalent is each?

\item \textbf{RQ2 (Distribution Across Components):}
  How are adaptations distributed across the specific components that make up a skill package?

\item \textbf{RQ3 (Family Co-occurrence):}
What specific adaptation families are most commonly bundled together during downstream adaptation?

\item \textbf{RQ4 (Security-Sensitive Additions):} What proportion of downstream adaptations add content matching security-sensitive patterns?

\item \textbf{RQ5 (Commit-Message Coverage):} To what extent do commit messages describe what changed and why it changed?

\end{itemize}

Our analysis exposes a fundamental mismatch between the intended design of autonomous agent skills and the reality of their evolution. While skills are published for reuse purpose, downstream engineers must spend significant effort to manually refit them by rewriting internal procedures, tightening constraints, altering tool/language targets, and forcing rediscoverability. 
We also find out that these overrides do not occur in isolation, changes to procedures, decisions, and policies are tightly coupled. 
Furthermore, downstream adaptations introduce an unexpected, invisible attack surface: nearly one-fifth (18.6\%) of forks actively inject security-sensitive patterns. Because these risks are embedded within natural-language instructions rather than code, they completely bypass traditional, code-focused security review pipelines. Finally, commit messages typically describe the functional outcome of a change rather than the underlying customization need, making direct inspection of the modification essential.


Our findings have implications for multiple stakeholders. Skill developers should provide clearer extension points, modular abstractions, explicit dependency structures, and standardized mechanisms for adapting tools, languages, constraints, and discovery metadata without rewriting core instructions.
There is a need for building tools to develop change-impact analysis, consistency checking, and adaptation-aware diffing techniques that can detect coupled modifications and surface security-sensitive changes in natural-language instructions. 
Finally, there are opportunities for researchers to investigate representations, testing methods, and security analyses tailored to evolving skill ecosystems, where behavioral logic is distributed across prompts, policies, metadata, and tool specifications.

We release our labeled corpus, taxonomy definitions, labeling workflow, audit
  materials, and analysis scripts to support future research on skill adaptation
  and skill ecosystems. 
We also release RADAR, a simple, lightweight, implication-driven review checklist that helps developers assess skill adaptations across different aspects.


\noindent This paper makes the following contributions:

\begin{enumerate}[leftmargin=*]

\item \textbf{First empirical study of downstream skill adaptation.}
  We construct a corpus of 1,126 skill-adaptation instances from six skill 
  repositories, providing the first empirical characterization of
  how downstream developers adapt published skills during reuse.

\item \textbf{A validated taxonomy of adaptation patterns.}
  We develop and validate a taxonomy comprising 46 adaptation patterns grouped
  into 13 families, capturing recurring forms of skill adaptation observed
  across independent downstream repositories.

\item \textbf{Open research artifacts.}
  We release the labeled corpus, taxonomy definitions, labeling workflow, audit
  records, and analysis scripts to support future research on skill adaptation
  and skill ecosystems, available on
  Section~\ref{sec:data-availability}.
\end{enumerate}

\section{Background and Related Work}\label{sec:background}

\subsection{LLM Agents}\label{sec:background-llm-agents}

An LLM agent is a system driven by an LLM that carries out multi-step
work by combining customized instructions, context, and
tool use.
The standard interaction paradigm interleaves reasoning and acting:
the agent reasons over the current task state, selects an action,
executes it through a tool or interface, observes the result, and
uses that observation as context for the next
step~\cite{yaoREACSYNERGIZINGREASONING2023}.
In software engineering, this pattern allows agents such as SWE-agent
and OpenHands to inspect files, edit code, run commands, and respond
to tool feedback within the same task
session~\cite{yangSWEagentAgentComputerInterfaces,
wangOPENHANDSOPENPLATFORM2025a}.

Related work on LLM agents has examined how agents access external
environments and knowledge rather than relying only on internal
parameters or task prompts.
Surveys of externalization describe memory, skills, protocols, and
harness engineering as mechanisms for giving agents reusable
context~\cite{zhouExternalizationLLMAgents2026}.
Skills sit in this design space as reusable packages that provide
procedures, guidance, and supporting resources for specific tasks
that agents can load when needed.

\subsection{The Agent Skills Ecosystem}
\label{sec:background-agent-skills}

\begin{figure}[t]
\centering
\makebox[\columnwidth][c]{%
\includegraphics[width=1.00\columnwidth]{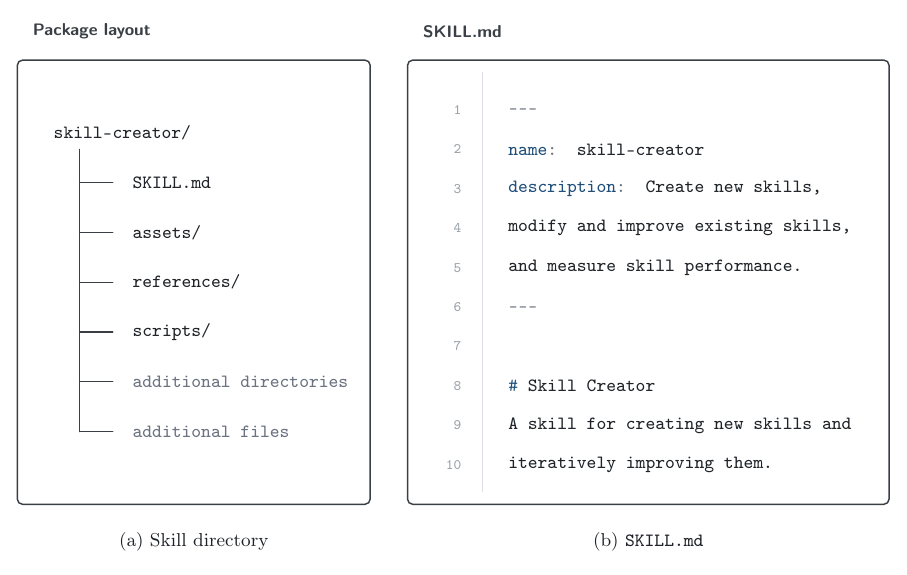}}
\vspace{-0.8cm}
\caption{A skill is a package directory
with a required \breaktt{SKILL.md} file. That file contains metadata,
including the required \breaktt{name} and routing
\breaktt{description}, together with task instructions for the agent.
Skills can also include bundled resources, which consist of files and
directories packaged with the skill. Individual files or directories
within the package are referred to as skill components. This shortened
\breaktt{skill-creator} example from \texttt{anthropics/skills}
collapses less relevant bundled resources into additional directories or files.}
\label{fig:skill-example}
\end{figure}

Skills (also known as agent skills) are reusable packages for LLM agents
that support particular tasks. Each skill is organized around a required
\texttt{SKILL.md} file and optional bundled resources, namely additional
files or directories such as scripts, templates, and reference
documents~\cite{agentskills}. Each \texttt{SKILL.md} file has two parts:
a YAML front matter that declares required metadata fields including
\texttt{name} and \texttt{description}, where \texttt{description} serves
as the routing signal used by an LLM agent to determine when the skill is
relevant; and a Markdown body that contains the instructions followed after
activation, including task procedures, decision rules, and references to
bundled resources. Within the Markdown body, headings and coherent
instruction units form body sections. We refer to each individual file or
directory within the skill package as a skill component.
Figure~\ref{fig:skill-example} shows this package structure.

LLM agents use skills through progressive disclosure, a staged loading
mechanism that controls which parts of a skill enter the agent's
working context~\cite{agentskills}.
During discovery, the agent reads only the metadata of available
skills.
When a task matches a skill's \texttt{description}, the agent
activates the skill and reads its body.
During execution, the agent reads or runs bundled resources only
when the body directs it to do so.
This mechanism matters for our study because metadata, body sections,
and bundled resources are distinct locations where a skill can be
modified.

Most related work on agent skills treats skills as reusable artifacts to
represent, acquire, retrieve, select, and evolve~\cite{zhouComprehensiveSurveyAgent2026a}.
Skill acquisition examines how skills are created from human expertise, agent
experience, task requirements, or external
resources~\cite{xuAgentSkillsLarge2026,alzubiEvoSkillAutomatedSkill2026,
yangAutoSkillExperienceDrivenLifelong2026,linMUSEAutoskillSelfEvolvingAgents2026,
gong2026skillmoo}, while skill retrieval and selection asks how an LLM agent
should discover, route, and compose skills from large
pools~\cite{jiangSoKAgenticSkills2026,zhengSkillRouterSkillRouting2026,
liOrganizingOrchestratingBenchmarking2026a}. Skill evolution concerns how
published skills are evaluated, updated, validated, and maintained over
time~\cite{dingAgentSkillEvaluation2026}. Related work also evaluates skill
effectiveness on downstream tasks and identifies security risks in shared
repositories~\cite{hanSWESkillsBenchAgentSkills2026,liSkillsBenchBenchmarkingHow2026,
liuAgentSkillsWild2026,liNoAttackRequired2026}. These studies establish the
importance of skills as reusable agent artifacts, but none characterizes how
downstream developers modify published skills in forked repositories.

\subsection{Mining Software Repositories (MSR)}\label{sec:background-msr}

MSR studies artifacts produced during software development, collected
from version control systems, issue trackers, code review systems,
and related development platforms~\cite{hassanRoadAheadMining2008,
gong2026analyzing}.
Because skills are stored as ordinary repository contents, their
changes are amenable to the same version-history perspective.
The GitHub fork model is especially relevant: developers derive a
public fork from an upstream repository, modify it independently,
and optionally contribute changes back, and prior MSR work has used
fork networks to characterize downstream activity and its relationship
to upstream projects~\cite{jiangWhyHowDevelopers2017}.
Mining GitHub data requires care because of visibility, activity,
and platform-specific biases~\cite{kalliamvakou2014promises}, and
changes on non-default branches may be missed by analyses limited
to the default branch alone~\cite{kovalenkoMiningFileHistories2018}.
Building on this line of MSR work, our study shifts the focus from
code changes in forked repositories to modifications of skill
components.

The closest prior work studies prompts in software repositories,
covering prompt patterns~\cite{whitePromptPatternCatalog2023,
whiteChatGPTPromptPatterns2023}, prompt evolution, and prompt
management~\cite{tafreshipourPromptingWildEmpirical,
liUnderstandingPromptManagement2026}.
Because skills are structured packages with metadata, instruction
bodies, and bundled resources rather than plain prompts, their
adaptations require a different analytic vocabulary than prompt
studies provide.

\section{Study Design}\label{sec:design}

\begin{figure}[t]
  \centering
  \includegraphics[width=\columnwidth]{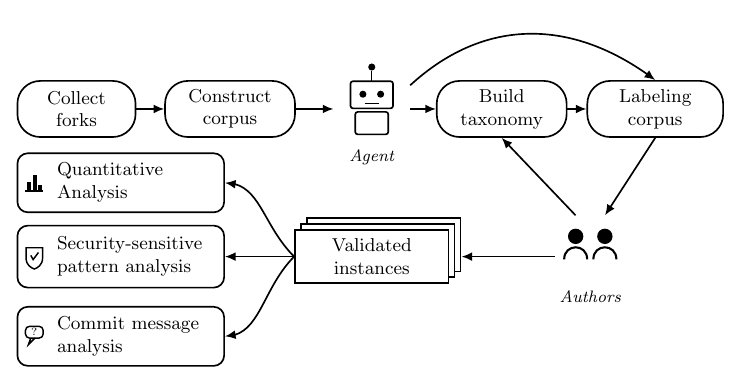}
  \vspace{-0.9cm}
  \caption{Overview of the study pipeline.}
  \label{fig:overview-pipeline}
\end{figure}

This study characterizes skill adaptation by comparing each branch in a
public fork against the default branch of its upstream repository and
classifying the observed modifications. We define an adaptation as a
modification that changes a skill package already present in the upstream
repository, and therefore exclude whole-skill additions or deletions,
upstream synchronization, repository maintenance changes, and files outside
skill packages. As Figure~\ref{fig:overview-pipeline} shows, the study
proceeds through five stages: corpus construction, taxonomy building, corpus
labeling, validation, and quantitative analysis; each stage feeds directly
into the next.

\subsection{Corpus Construction}\label{sec:corpus-construction}

We collected our corpus from six upstream repositories dedicated to
reusable agent skills, listed in Table~\ref{tab:upstream}. Public fork
histories provide observable traces of downstream skill adaptation, a
framing consistent with prior MSR work that treats forks as evidence of
software reuse and divergence~\cite{hassanRoadAheadMining2008,jiangWhyHowDevelopers2017}.
We therefore used these upstream repositories as seeds for enumerating
downstream forks and their branch histories.

Candidate upstream repositories were identified by searching GitHub on
2026-04-28 with the keywords \texttt{skill} and \texttt{agent skill},
requiring at least 20,000 stars and 2,000 forks, and sorting results by
stars~\cite{github_rest_api_docs}; the two result sets were merged and
deduplicated by GitHub full name. We then screened the deduplicated
candidates by traversing each repository's default-branch file tree,
retaining only repositories that contained at least one \texttt{SKILL.md}
file, and manually excluding repositories where skills served only as
auxiliary support for a broader project rather than as primary reusable
artifacts. The exact search query, GitHub API parameters, and pagination
logic are released as the search script in the replication artifact
(Section~\ref{sec:data-availability}). Because the study relies on public
fork histories as evidence of downstream adaptation, we applied star and
fork thresholds to ensure sufficient downstream fork activity; the six
selected repositories therefore serve as upstream seeds for enumerating
downstream forks rather than as a representative sample of all skill
repositories~\cite{kalliamvakou2014promises}.


\begin{table}[t]
  \centering
  \caption{Selected upstream repositories. Stars and forks are GitHub
  counts captured on 2026-04-28.}
  \vspace{-0.3cm}
  \label{tab:upstream}
  \resizebox{0.8\columnwidth}{!}{%
  \begin{tabular}{@{}lrr@{}}
    \toprule
    Repository & Stars & Forks \\
    \midrule
    \texttt{anthropics/skills} & 125,742 & 14,697 \\
    \texttt{obra/superpowers} & 172,158 & 15,025 \\
    \texttt{affaan-m/everything-claude-code} & 169,700 & 26,239 \\
    \texttt{mattpocock/skills} & 40,040 & 2,764 \\
    \texttt{vercel-labs/agent-skills} & 25,892 & 2,353 \\
    \texttt{ComposioHQ/awesome-claude-skills} & 56,990 & 6,186 \\
    \bottomrule
  \end{tabular}
  }
\end{table}


\begin{table}[t]
  \centering
  \caption{Corpus construction and filtering process from public forks to branch comparison records. Counts are computed from data collected on 2026-04-28.}
  \vspace{-0.3cm}
  \label{tab:filtering-process}
  \resizebox{\columnwidth}{!}{%
  \begin{tabular}{@{}llr@{}}
    \toprule
    Construction step & Unit & Count \\
    \midrule
    Enumerate public forks & forks & 67,264 \\
    Remove trivial forks & forks & 6,516 \\
    Compare branches and remove shared history & branch comparisons & 3,266 \\
    Retain in-skill changes & branch comparisons & 1,451 \\
    Deduplicate redundant branches & branch comparisons & 1,220 \\
    \bottomrule
  \end{tabular}
  }
\end{table}

For each selected repository, we enumerated all public forks through the
GitHub API and removed forks whose latest push timestamp was no more than
five minutes after their creation timestamp, treating such forks as inactive
and unlikely to contain downstream adaptation. This heuristic reduced the
67,264 enumerated forks to 6,516 retained candidates.


Forks inherit the commit history of their upstream repositories. A naive
branch comparison therefore returns downstream changes introduced
by the fork owner as well as commits inherited from upstream or shared across forks. For each retained fork, we compared every available branch
against the upstream default branch using the GitHub Compare
API~\cite{github_rest_api_docs}, rather than only the fork's default branch,
to avoid missing modifications stored in feature or maintenance
branches~\cite{kovalenkoMiningFileHistories2018}. Each comparison produced candidate downstream commits, changed files, and merge-base commit
distinguishing the fork branch from the upstream base. We then removed
commits whose SHAs appeared under any other fork owner of the same upstream
repository, retaining only commits unique to each owner and excluding
comparisons with no remaining commits, yielding 3,266 branch comparison
records.

To focus the corpus on adaptations of existing skill packages, we retained
only branch comparisons that modified files inside skill packages already
present on the upstream default branch, defined as directories containing a
\texttt{SKILL.md} file. This step excluded skill additions and
deletions, binary artifacts, and repository changes unrelated to existing
skill packages. We additionally removed comparisons whose commit subjects
indicated upstream synchronization rather than downstream adaptation,
reducing the 3,266 branch comparison records to 1,451.

Before labeling, we cleaned commit messages by removing mechanical subjects
such as merges, squashes, and reverts; repository maintenance subjects
including CI and dependency updates; and auto-generated footer lines. We
then deduplicated branches within each fork owner by collapsing identical
commit sets into one representative record and replacing subset branches
with the branch carrying the larger commit set, reducing the 1,451 branch
comparison records to 1,220.

\subsection{Taxonomy Building}\label{sec:taxonomy-building}

To identify recurring adaptation patterns, we constructed an initial taxonomy
through an inductive labeling process on a stratified sample, using an LLM
as the labeling agent, following recent SE studies that apply LLMs to
qualitative labeling~\cite{ahmedCanLLMsReplace2025,wangCanLLMsReplace2025}.
The labeling agent was \texttt{claude-opus-4-8} via the \texttt{Claude Code}
CLI configured with medium reasoning effort, a setting that balances
labeling quality, consistency, and token cost at corpus
scale~\cite{anthropic_effort}. The taxonomy-building prompts, structured
output schema, and per-pattern decision rules are released with the
replication artifact (Section~\ref{sec:data-availability}).

We sampled 300 of the 1,220 branch comparison records using a fixed random
seed and stratifying by upstream repository. For each record, the labeling
agent received the patch contents, cleaned commit messages, and file-level
metadata, then produced candidate entries for a two-level taxonomy: at the
first level, each \emph{family} captured a broad adaptation purpose; at the
second level, each \emph{pattern} belonged to one family and specified a
name, definition, patch-evidence requirement, and decision rule.

Two authors reviewed the candidate taxonomy, checking that each family
represented a distinct adaptation purpose, each pattern was grounded in
observable patch evidence, and decision rules separated adjacent patterns;
merging redundant patterns, refining ambiguous definitions, and removing
categories that reflected repository maintenance rather than skill adaptation.

We then expanded the taxonomy iteratively over the remaining 920 records in
batches of 50. For each batch, the labeling agent labeled each record
against the current taxonomy; when no existing pattern adequately described
a record, we proposed a candidate pattern or family following the same
schema, which two authors reviewed before adding it to the taxonomy. The
taxonomy was considered saturated when two consecutive batches introduced no
new pattern or family.

\subsection{Labeling Corpus}\label{sec:labeling-corpus}

Using the saturated taxonomy as a fixed label vocabulary, we labeled all
1,220 branch comparison records with the same labeling agent as in
Section~\ref{sec:taxonomy-building}. For each record, the agent
received the patch contents, cleaned commit messages, file metadata, and the
full taxonomy, then evaluated each pattern's decision rule and required
patch evidence before assigning labels. Multiple labels per record were
permitted, with the most specific applicable pattern preferred; records for
which no pattern applied received an empty label set with a stated rationale
rather than a forced label.

Each assigned label included supporting evidence and a confidence level
(\textit{high}, \textit{medium}, or \textit{low}), used to identify records
for mandatory inclusion in the validation audit~\cite{wangHumanLLMCollaborativeAnnotation2024}.
\textit{High} confidence required direct patch evidence with minimal
ambiguity; \textit{medium} confidence indicated partially ambiguous but
sufficient evidence; \textit{low} confidence flagged indirect, incomplete,
or commit-message-derived evidence.

\subsection{Validation}
\label{sec:validation}

After labeling, we conducted a label verification audit of all 1,220 branch
comparison records. We drew a random sample of 293 records using Cochran's
formula with finite population correction~\cite{cochran1977sampling} at
95\% confidence and a 5\% margin of error. Following prior work on human
verification of LLM labels~\cite{wangHumanLLMCollaborativeAnnotation2024},
we additionally included all 35 records carrying at least one low-confidence
label. Low-confidence labels indicate weak or indirect evidence and a higher
risk of error, so all such records were included for human review. Of these
35, 10 were already in the random sample and the remaining 25 were added,
yielding 318 records for audit. For each record, two authors independently
reviewed the agent-assigned labels, supporting rationale, patch contents,
and commit messages, recording the additions and removals needed to correct
the label set.

We report two symmetric agreement comparisons: between the two independent
auditors to assess the reliability of the human reference, and between the
consensus and the original agent labels to measure labeling accuracy. The
two authors reconciled disagreements through
discussion~\cite{wangHumanLLMCollaborativeAnnotation2024}, removing labels
unsupported by patch evidence and removing label sets of records whose diffs
carried no inspectable content. Since neither the auditor labels nor the
agent labels serve as the ground truth, we report symmetric agreement
metrics rather than precision and recall: positive agreement, Jaccard,
Cohen's $\kappa$~\cite{cohen1960coefficient}, and Krippendorff's
$\alpha$~\cite{krippendorff2004reliability}, each computed per label as
binary present/absent decisions. Table~\ref{tab:validation-metrics} reports
both comparisons.

\begin{table}[t]
  \centering
  \caption{Validation metrics for the 318 audited records.}
  \vspace{-0.3cm}
  \label{tab:validation-metrics}
  \resizebox{\columnwidth}{!}{%
  \begin{tabular}{@{}lrrrrr@{}}
    \toprule
    Comparison & Items & Pos.\ agr. & Jaccard & Cohen's \(\kappa\) & Krippendorff \(\alpha\) \\
    \midrule
    Auditor A vs.\ B & 318 & 0.8795 & 0.8469 & 0.8323 & 0.8722 \\
    Consensus vs.\ Agent & 318 & 0.8837 & 0.7849 & 0.8471 & 0.8767 \\
    \bottomrule
  \end{tabular}
  }
\end{table}

The validation results confirm both the reliability of the human consensus
and the quality of the agent labels. The two auditors reached strong
agreement after chance correction ($\kappa$=0.83, $\alpha$=0.87), and
the agent labels closely matched the reconciled consensus ($\kappa$=0.85,
$\alpha$=0.88).

Reconciliation moved in opposite directions across the two audit sets. On
the 293-record random sample, auditors more often added labels the agent had
omitted (136 added, 30 removed), indicating that the agent labeled
conservatively on typical records. On the 25 low-confidence records, auditors
almost exclusively removed unsupported labels (25 removed, 2 added),
confirming that the low-confidence flag effectively isolates labels assigned
without sufficient evidence: 31 of the 38 low-confidence labels (81.6\%) were
removed during reconciliation, and 20 of the 25 added records received no
label after reconciliation.

The final analysis corpus treats each branch comparison record with at least
one validated adaptation label as one skill-adaptation instance. Audited
records use the reconciled consensus labels; unaudited records retain the
original agent labels. This yielded 1,126 skill-adaptation instances: 277
from audited records and 849 from unaudited records.

To bound the reliability of the 849 unaudited records, we measured
agent labels against the reconciled consensus on the random sample,
which is an unbiased draw from the full corpus (the low-confidence
set is excluded because it is oversampled).
At the label level, agent precision is 95.9\% (95\% Wilson interval
[94.2, 97.1]) and recall is 83.9\% ([81.2, 86.2]).
Because the consensus was reconciled from the agent output, precision
is an upper bound rather than an independent estimate; the
inter-auditor agreement ($\kappa$=0.83) reported above is the
independent reliability estimate.
The residual error is asymmetric: roughly 4\% of agent labels are
removed under audit, so unaudited prevalences are slight
overestimates, while the larger missed-label rate mainly lowers
recall for secondary labels.
We therefore report Wilson intervals for headline prevalences so
that point estimates carry explicit uncertainty bounds (for example,
lifecycle 40.9\% [38.1, 43.8] and the security-sensitive share
18.6\% [16.4, 20.9]).

\subsection{Quantitative Analysis}
\label{sec:analysis}

The retained instances are unevenly distributed across upstream repositories
(Table~\ref{tab:instance-coverage}), so we report four metrics: branch
prevalence as the primary estimate, skill breadth and upstream coverage as
spread indicators, and repository-adjusted prevalence as a sensitivity check
for repository imbalance.

\begin{table}[t]
\centering
\caption{Distribution of retained skill-adaptation instances by upstream
repository.}
\vspace{-0.3cm}
\label{tab:instance-coverage}
\resizebox{0.85\columnwidth}{!}{%
\begin{tabular}{@{}lrr@{}}
\toprule
Upstream repository & Instances & Share (\%) \\
\midrule
\breaktt{obra/superpowers} & 694 & 61.6 \\
\breaktt{anthropics/skills} & 204 & 18.1 \\
\breaktt{affaan-m/everything-claude-code} & 155 & 13.8 \\
\breaktt{vercel-labs/agent-skills} & 47 & 4.2 \\
\breaktt{ComposioHQ/awesome-claude-skills} & 24 & 2.1 \\
\breaktt{mattpocock/skills} & 2 & 0.2 \\
\bottomrule
\end{tabular}
}
\end{table}

Branch prevalence measures the proportion of skill-adaptation instances
containing a given adaptation pattern:

\[
\mathrm{Prevalence}(p)=\frac{|I_p|}{|I|}
\]

where \(I_p\) is the set of instances containing pattern \(p\), and \(I\)
is the set of all skill-adaptation instances.

Skill breadth measures how widely a pattern appears across distinct skills:

\[
\mathrm{SkillBreadth}(p)=\frac{|S_p|}{|S|}
\]

where \(S_p\) is the set of skills in instances containing pattern \(p\),
and \(S\) is the set of all observed skills. Each skill is identified by its
upstream repository and root directory.

Upstream coverage measures the proportion of upstream repositories in which
a pattern appears:

\[
\mathrm{UpstreamCoverage}(p)=\frac{|U_p|}{|U|}
\]

where \(U_p\) is the set of upstream repositories containing pattern \(p\),
and \(U\) is the set of all upstream repositories.

Repository-adjusted prevalence reduces the influence of upstream repositories
that contribute a disproportionately large number of instances:

\[
\mathrm{AdjustedPrevalence}(p)=
\sum_{u \in U}
w_u
\frac{|I_{p,u}|}{|I_u|}
\]

where \(I_{p,u}\) is the set of instances from repository \(u\) containing
pattern \(p\), and \(I_u\) is the set of all instances from repository
\(u\). Repository weights are defined as:

\[
w_u=
\frac{\sqrt{|I_u|}}
{\sum_{v \in U}\sqrt{|I_v|}}
\]

Square-root weighting moderates the influence of large repositories while
retaining their relative contribution~\cite{gelman2007data}.

The same metrics apply to taxonomy families by aggregating the patterns
within each family, and to co-occurring families (bundles), defined as sets
of families that appear together in the same skill-adaptation instance.
For co-occurrence analysis, we additionally report lift, which measures
how much more often a set co-occurs than expected under independence of
its members.

\subsection{Security-Sensitive Pattern Analysis}
\label{sec:security-sensitive-analysis}

We identify security-sensitive modifications using the static rule-based
scan from the security pattern catalog proposed by Liu et
al.~\cite{liuAgentSkillsWild2026}, which covers seven categories relevant
to agent skills: external network requests, access to sensitive files,
dangerous file operations, command execution, dynamic or injected code
execution, dependency changes, and obfuscation. Each rule is a regular
expression over unified-diff content; a match flags a patch line as
potentially requiring security review. Table~\ref{tab:security-rule-patterns}
lists the rule patterns; descriptions are simplified, and the full regular
expressions and allow-list configuration are released with the replication
artifact.

We made one adjustment to the catalog. The sensitive-file rule matches the
bare word \texttt{token}, but in skill modification content this word
predominantly refers to language-model units such as context tokens and token
budgets rather than authentication credentials. We therefore restrict this
rule to credential-bearing forms such as \texttt{access\_token} and
\texttt{auth\_token} to avoid flagging benign language-model references.

\begin{table}[t]
  \centering
  \caption{Simplified security-sensitive rule patterns used by the static scan (adapted from~\cite{liuAgentSkillsWild2026}).}
  \vspace{-0.3cm}
  \label{tab:security-rule-patterns}
  \scriptsize
  \resizebox{\columnwidth}{!}{%
  \begin{tabular}{@{}l@{\hspace{1em}}p{0.55\columnwidth}@{}}
    \toprule
    Name & Description \\
    \midrule
    \rowcolor{gray!20}[0pt][0pt] \multicolumn{2}{@{}l@{}}{\textit{network}} \\
    External Network Requests & Non-allowed external requests \\
    Data Leakage & Outbound local data transfer \\
    \rowcolor{gray!20}[0pt][0pt] \addlinespace[1.2ex]
    \multicolumn{2}{@{}l@{}}{\textit{file}} \\
    Access to Sensitive Files & Secret or credential references \\
    Dangerous File Operations & Operations on protected paths \\
    \rowcolor{gray!20}[0pt][0pt] \addlinespace[1.2ex]
    \multicolumn{2}{@{}l@{}}{\textit{command}} \\
    Execution of Dangerous Commands & Destructive or privileged commands \\
    System Command Invocation & Shell or subprocess invocation \\
    \rowcolor{gray!20}[0pt][0pt] \addlinespace[1.2ex]
    \multicolumn{2}{@{}l@{}}{\textit{injection}} \\
    Code Injection & Code inserted into files \\
    Dynamic Code Execution & Dynamic runtime execution \\
    Backdoor Implantation & Reverse shell or backdoor logic \\
    \rowcolor{gray!20}[0pt][0pt] \addlinespace[1.2ex]
    \multicolumn{2}{@{}l@{}}{\textit{dependency}} \\
    Global Package Installation & Global package installs \\
    Forced Version Overrides & Forced dependency overrides \\
    \rowcolor{gray!20}[0pt][0pt] \addlinespace[1.2ex]
    \multicolumn{2}{@{}l@{}}{\textit{obfuscation}} \\
    Code Obfuscation & Encoded command execution \\
    Hidden Commands & Indirect command invocation \\
    \bottomrule
  \end{tabular}
  }
\end{table}

The analysis operates at the patch level, where each patch is the unified
diff of one skill-adaptation instance (Section~\ref{sec:validation}). We
focus on security-sensitive content introduced by downstream adaptations,
flagging a patch when any added line matches at least one rule. For each
flagged patch, we record the matched rule identifiers, pattern categories,
affected files, and matched line counts.

\subsection{Commit Message Analysis}
\label{sec:commit-analysis}

The commit-message analysis uses the 277 of the 318 audited records
(Section~\ref{sec:validation}) that retained at least one adaptation-pattern
label after consensus, as the \emph{What} analysis requires a reference
label set to compare against commit-message labels. Following Tian et
al.~\cite{tianWhatMakesGood2022}, we distinguish \emph{What}, referring to
adaptation content, from \emph{Why}, referring to the stated rationale
behind the modification. Table~\ref{tab:commit-message-taxonomy} summarizes
the commit-message expression categories used in this analysis.

\begin{table*}[t]
\centering
\caption{Commit-message expression categories adapted from Tian et al.~\cite{tianWhatMakesGood2022}.}
\label{tab:commit-message-taxonomy}
\vspace{-0.3cm}
\resizebox{\textwidth}{!}{%
\begin{tabular}{@{}lll@{}}
\toprule
Dimension & Category & Definition \\
\midrule

\multirow{4}{*}{\emph{What}}
& Summarize skill artifact change
& Names the affected skill artifact, such as a skill, file, resource, section, metadata field, script, or packaged component. \\
& Illustrate skill function
& Describes the affected skill behavior, capability, workflow, or user-facing functionality. \\
& Describe skill implementation principle
& Explains the technical mechanism, execution principle, or implementation strategy behind the modification. \\
& Missing What
& Provides no codeable information describing what changed. \\

\midrule

\multirow{5}{*}{\emph{Why}}
& Describe skill issue
& Motivates the modification by describing a bug, failure, limitation, inconsistency, or other issue affecting the skill or its supporting artifacts. \\
& Illustrate skill requirement
& Motivates the modification by describing a requirement, such as an upstream dependency, platform constraint, workflow need, or external integration requirement. \\
& Describe skill objective
& Explains the intended objective of the modification, such as introducing a new capability, improving quality, or achieving an expected outcome. \\
& Imply skill necessity
& Indicates that the modification is necessary without explicitly stating an issue, requirement, or objective. \\
& Missing Why
& Provides no codeable rationale explaining why the modification was made. \\

\bottomrule
\end{tabular}%
}
\end{table*}

Commit-message labeling followed the same agent-then-audit workflow as
Section~\ref{sec:labeling-corpus}, with one difference: the labeling agent
received only the cleaned commit messages, with patch contents, file
metadata, and reconciled patch labels withheld. Two authors then
independently audited the agent-assigned labels using the same
message-only evidence and reconciled disagreements through discussion. The
final labels therefore reflect what can be inferred from commit messages
alone, independent of patch content.

To evaluate \emph{What}, we compare commit-message labels against the
reconciled patch labels for the same 277 records and report two measures:
\emph{recall}, the average proportion of patch labels captured by the
corresponding commit-message label set; and \emph{any-hit rate}, the
percentage of records whose commit-message labels capture at least one patch
label. Both measures are reported at the pattern and family levels.

For \emph{Why}, we report the percentage of records containing at least one
non-missing rationale category and the distribution across rationale
categories. We additionally examine whether adaptation families exhibit
distinct rationale profiles.












\section{Results}\label{sec:results}

\subsection{RQ1. Adaptation Patterns and Prevalence}

\subsubsection{RQ1a. What patterns and families characterize skill
adaptation?}


\definecolor{ink}{HTML}{1E466E}

\definecolor{cLifecycle}{HTML}{4E659B}
\definecolor{cRetarget}{HTML}{8AA3B0}
\definecolor{cProcedure}{HTML}{72BCD5}
\definecolor{cDecision}{HTML}{E76254}
\definecolor{cPolicy}{HTML}{EF8A47}
\definecolor{cGuardrail}{HTML}{1E466E}
\definecolor{cSpec}{HTML}{FFD06F}
\definecolor{cStyle}{HTML}{B8A8CF}
\definecolor{cPolish}{HTML}{AADCE0}
\definecolor{cResource}{HTML}{528FAD}
\definecolor{cPersonalize}{HTML}{B6766C}
\definecolor{cScript}{HTML}{376795}
\definecolor{cFix}{HTML}{E58B7B}

\newcommand{\curfam}{ink}
\newif\iffirstfam \firstfamtrue

\newcommand{\famhdr}[4]{%
  \iffirstfam\global\firstfamfalse\else\noalign{\vskip 5pt}\fi
  \gdef\curfam{#1}%
  \cellcolor{#1} &
  \cellcolor{#1!16}{\sffamily\bfseries\small\textcolor{#1!60!black}{#2}%
     \hspace{0.4em}\textcolor{#1!55!black}{\sffamily\scriptsize\mdseries#3}} &
  \cellcolor{#1!16}{\sffamily\scriptsize\itshape\textcolor{black!68}{#4}} \\
}

\newcommand{\lbl}[2]{%
  \cellcolor{\curfam} &
  {\ttfamily\scriptsize\textcolor{ink}{#1}} &
  {\footnotesize\textcolor{black!82}{#2}} \\
}

\begin{table*}[t]
  \centering
  \caption{\textbf{Adaptation taxonomy: 46 adaptation patterns across 13 families.}}
  \vspace{-0.3cm}
  \label{tab:categorization}
  \resizebox{\textwidth}{!}{%

\begin{minipage}{24.2cm}
\setlength{\tabcolsep}{5pt}

\begin{minipage}[t]{11.9cm}
\firstfamtrue
\renewcommand{\arraystretch}{1.32}
\begin{tabular}{@{}p{0.17cm}@{\hspace{11pt}}>{\raggedright\arraybackslash}p{4.65cm}>{\raggedright\arraybackslash}p{6.25cm}@{}}

\famhdr{cLifecycle}{LIFECYCLE}{5 patterns}{Package and present skills.}
\lbl{rename-or-reorganize-structure}{Rename, move, or reorganize files.}
\lbl{remove-auxiliary-file}{Delete auxiliary files.}
\lbl{translate-skill-content}{Add or maintain translations.}
\lbl{modify-skill-metadata}{Edit frontmatter or metadata.}
\lbl{style-or-format-content}{Fix formatting or grammar.}

\famhdr{cRetarget}{RETARGET}{3 patterns}{Adapt technology or host.}
\lbl{retarget-domain-stack}{Change language, stack, or domain.}
\lbl{retarget-version-api}{Update version or API guidance.}
\lbl{retarget-host-tooling}{Adapt to another host or tool.}

\famhdr{cProcedure}{PROCEDURE}{9 patterns}{Change workflow steps.}
\lbl{add-procedure-step}{Add procedure steps.}
\lbl{add-sub-workflow}{Add a named multi-step sub-workflow.}
\lbl{remove-procedure-step}{Remove steps or sub-workflows.}
\lbl{add-review-round}{Add review or self-check rounds.}
\lbl{reorder-steps}{Reorder, split, or merge steps.}
\lbl{add-platform-specific-branch}{Add platform- or client-specific branches.}
\lbl{swap-execution-model}{Change delegation model.}
\lbl{integrate-external-system}{Route through an external system.}
\lbl{change-artifact-storage}{Change artifact storage or names.}

\famhdr{cDecision}{DECISION}{4 patterns}{Control activation and path choice.}
\lbl{add-activation-criterion}{Add activation criteria.}
\lbl{add-decision-rule}{Add branching or dispatch rules.}
\lbl{add-consent-gate}{Require user confirmation.}
\lbl{add-precondition-check}{Add pre-action checks.}

\famhdr{cGuardrail}{GUARDRAIL}{1 pattern}{Add instruction safety protections.}
\lbl{add-instruction-guardrail}{Add safety or security guardrails.}

\famhdr{cPolish}{POLISH}{1 pattern}{Sharpen instruction wording.}
\lbl{tighten-instruction-wording}{Clarify wording without new rules.}

\end{tabular}
\end{minipage}%
\hspace{0.4cm}%
\begin{minipage}[t]{11.9cm}
\firstfamtrue
\renewcommand{\arraystretch}{1.32}
\begin{tabular}{@{}p{0.17cm}@{\hspace{11pt}}>{\raggedright\arraybackslash}p{4.65cm}>{\raggedright\arraybackslash}p{6.25cm}@{}}

\famhdr{cPolicy}{POLICY}{4 patterns}{Set behavioral rules.}
\lbl{add-priority-hierarchy}{Set rule or source precedence.}
\lbl{add-design-principle}{Add design or method principles.}
\lbl{add-hard-constraint}{Add mandatory or prohibitive rules.}
\lbl{relax-constraint}{Weaken or remove constraints.}

\famhdr{cSpec}{SPEC}{4 patterns}{Specify outputs and artifacts.}
\lbl{add-output-format-spec}{Define output structure or fields.}
\lbl{add-scoring-rubric}{Add criteria, anchors, or scales.}
\lbl{add-example}{Add examples, samples, or worked cases.}
\lbl{add-template}{Add reusable templates or skeletons.}

\famhdr{cStyle}{STYLE}{1 pattern}{Add writing-style guidance.}
\lbl{add-voice-rule}{Add tone, register, or style rules.}

\famhdr{cResource}{RESOURCE}{5 patterns}{Add bundled local files.}
\lbl{add-reference-doc}{Add local reference documents.}
\lbl{add-prompt-template-resource}{Add prompt template resources.}
\lbl{add-tool-config}{Add tool or provider configuration.}
\lbl{add-eval-suite-resource}{Add tests, fixtures, or eval assets.}
\lbl{add-output-asset}{Add generated artifacts or showcases.}

\famhdr{cPersonalize}{PERSONALIZE}{1 pattern}{Layer fork-owner defaults.}
\lbl{personalize-defaults}{Add fork-owner preferences or defaults.}

\famhdr{cScript}{SCRIPT}{7 patterns}{Modify local executable scripts.}
\lbl{script-add-feature}{Add script features, flags, or modes.}
\lbl{script-add-safety-guard}{Add script safety checks.}
\lbl{script-add-input-validation}{Validate inputs, paths, or schemas.}
\lbl{script-optimize}{Improve performance or token cost.}
\lbl{script-fix-logic-bug}{Fix script logic defects.}
\lbl{script-fix-platform-compat}{Fix OS, shell, or library compatibility.}
\lbl{script-refactor}{Restructure scripts without behavior change.}

\famhdr{cFix}{FIX}{1 pattern}{Correct broken instructions.}
\lbl{fix-skill-instruction-bug}{Fix broken, stale, or misleading instructions.}

\end{tabular}
\end{minipage}

\end{minipage}

  }
\end{table*}

\arrayrulecolor{black}

Table~\ref*{tab:categorization} presents the 46 adaptation patterns and 13
families identified through the taxonomy building process
(Section~\ref{sec:taxonomy-building}) and applied to the 1,126
skill-adaptation instances from six upstream repositories
(Section~\ref{sec:analysis}). Each pattern is listed under its family with a
brief description. The families are organized by the part of a skill they
adapt, following its layered package structure
(Section~\ref{sec:background-agent-skills}): eight families (procedure,
decision, policy, spec, guardrail, polish, fix, and personalize) modify the
natural-language instructions in the body that govern agent behavior, while the
remaining five act outside the body, covering packaging and presentation
(lifecycle), technology or host retargeting (retarget), bundled local files
(resource), executable scripts (script), and writing style (style). The
taxonomy thus spans both the prose an agent reads and the structural and
executable surfaces around it.

\subsubsection{RQ1b. How prevalent are the identified patterns and families?}

\begin{table}[t]
  \centering
  \caption{Top 10 adaptation patterns ranked by branch prevalence.}
  \vspace{-0.3cm}
  \label{tab:rq1-pattern-metrics}
  \resizebox{\columnwidth}{!}{%
  \begin{tabular}{@{}lrrrr@{}}
    \toprule
    Pattern & Branch prev.(\%) & Skill brd.(\%) & Upstream cov. & Adjusted prev.(\%) \\
    \midrule
    \breaktt{modify-skill-metadata} & 18.2 & 77.8 & 5/6 & 19.6 \\
    \breaktt{add-procedure-step} & 17.8 & 7.5 & 5/6 & 12.3 \\
    \breaktt{add-decision-rule} & 15.1 & 4.9 & 4/6 & 10.8 \\
    \breaktt{add-hard-constraint} & 14.6 & 4.1 & 5/6 & 10.6 \\
    \breaktt{integrate-external-system} & 12.1 & 4.0 & 3/6 & 8.1 \\
    \breaktt{retarget-host-tooling} & 11.5 & 12.9 & 4/6 & 8.6 \\
    \breaktt{translate-skill-content} & 11.5 & 28.7 & 5/6 & 13.8 \\
    \breaktt{add-reference-doc} & 10.9 & 11.2 & 5/6 & 10.5 \\
    \breaktt{change-artifact-storage} & 10.5 & 2.7 & 3/6 & 7.0 \\
    \breaktt{relax-constraint} & 9.2 & 3.3 & 3/6 & 6.3 \\
    \bottomrule
  \end{tabular}
  }
\end{table}

\begin{table}[t]
  \centering
  \caption{All 13 adaptation families ranked by branch prevalence.}
  \vspace{-0.3cm}
  \label{tab:rq1-family-metrics}
  \resizebox{\columnwidth}{!}{%
  \begin{tabular}{@{}lrrrr@{}}
    \toprule
    Family & Branch prev.(\%) & Skill brd.(\%) & Upstream cov. & Adjusted prev.(\%) \\
    \midrule
    lifecycle & 40.9 & 91.4 & 5/6 & 42.7 \\
    procedure & 40.8 & 12.7 & 5/6 & 29.3 \\
    decision & 25.3 & 9.6 & 5/6 & 18.3 \\
    policy & 23.9 & 7.0 & 5/6 & 17.4 \\
    resource & 16.8 & 24.7 & 5/6 & 16.2 \\
    retarget & 16.3 & 15.1 & 5/6 & 14.6 \\
    script & 16.2 & 24.7 & 5/6 & 19.4 \\
    spec & 11.8 & 6.8 & 5/6 & 10.0 \\
    fix & 8.5 & 10.8 & 5/6 & 9.0 \\
    polish & 4.0 & 11.2 & 5/6 & 3.6 \\
    personalize & 3.4 & 11.0 & 5/6 & 5.7 \\
    guardrail & 1.9 & 3.1 & 4/6 & 1.8 \\
    style & 0.5 & 2.4 & 2/6 & 0.4 \\
    \bottomrule
  \end{tabular}
  }
\end{table}

The prevalence distribution in Tables~\ref{tab:rq1-pattern-metrics}
and~\ref{tab:rq1-family-metrics} reveals three recurring needs. The most
prevalent pattern, \breaktt{modify-skill-metadata}, has the widest spread of
any pattern; 85.4\% of its instances change the \texttt{description} field,
which serves as the routing signal for skill activation, indicating a
recurring need to make a skill reliably discoverable before any behavioral
change is made. The next most prevalent patterns all rewrite the
instructional body, \breaktt{add-procedure-step}, \breaktt{add-decision-rule},
and \breaktt{add-hard-constraint}, aggregating into the procedure, decision,
and policy families; unlike metadata editing, these are concentrated in a
narrow set of skills yet remain the leading instructional adaptations even
after repository adjustment. Finally, \breaktt{retarget-host-tooling} and
\breaktt{translate-skill-content} aggregate into the retarget and lifecycle
families, porting a skill's behavior to a different host tool, technology
stack, or natural language; the latter alone spans over a quarter of all
skills, indicating that localization recurs broadly.

Lifecycle is the single most prevalent and most widely spread family,
robust to repository adjustment, while guardrail and style reflect
occasional rather than systematic adaptation.

\begin{rqanswer}{RQ1}
    We identify 46 adaptation patterns organized into 13 families. 
    Metadata, procedure, and policy adaptations are the most prevalent.
    Their dominance exposes a reuse paradox: although skills are published for plug-and-play adoption, developers must still rewrite how they are discovered and executed to fit local contexts.
\end{rqanswer}

\subsection{RQ2. Component Distribution}

\begin{table}[t]
  \centering
  \caption{Distribution of modifications across skill directory components.}
  \vspace{-0.3cm}
  \label{tab:rq2-components}
  \resizebox{\columnwidth}{!}{%
  \begin{tabular}{@{}lrrrr@{}}
    \toprule
    Component & Branch prev.(\%) & Skill brd.(\%) & Upstream cov. & Adjusted prev.(\%) \\
    \midrule
    \texttt{SKILL.md} & 79.8 & 75.9 & 6/6 & 72.9 \\
    additional files & 40.9 & 43.5 & 5/6 & 36.2 \\
    scripts & 23.3 & 30.9 & 5/6 & 25.9 \\
    references & 15.0 & 27.7 & 5/6 & 14.1 \\
    examples & 8.4 & 10.2 & 4/6 & 6.6 \\
    configuration & 4.4 & 41.0 & 5/6 & 6.7 \\
    additional directories & 2.7 & 7.4 & 3/6 & 2.6 \\
    assets & 2.2 & 10.1 & 3/6 & 2.8 \\
    \bottomrule
  \end{tabular}
  }
\end{table}

\texttt{SKILL.md} is modified in 79.8\% of instances (Table~\ref{tab:rq2-components}), the only component
touched across all six upstream repositories, and remains dominant after
repository adjustment (72.9\%). Scripts (23.3\%) and references (15.0\%) are
secondary surfaces, each present across five repositories. Configuration is
modified in only 4.4\% of instances yet present in 41.0\% of observed skills,
indicating that downstream developers adapt what a skill instructs more than
how it is configured.

\begin{rqanswer}{RQ2}
    With 79.8\% of modifications targeting \texttt{SKILL.md} and only 23.3\% targeting code scripts, developers have clearly established the Markdown file as the agent's absolute control plane, demoting traditional executable code to a secondary, supporting role.
\end{rqanswer}

\subsection{RQ3. Family Co-occurrence}

\begin{table}[t]
\centering
\caption{Most common adaptation bundles ranked by branch prevalence. Lift
measures the association strength between adaptation families after
accounting for their individual frequencies.}
\vspace{-0.3cm}
\label{tab:rq3-family-bundles}
\resizebox{\columnwidth}{!}{%
\begin{tabular}{@{}lrrrr@{}}
\toprule
Bundle & Branch prev.(\%) & Skill brd.(\%) & Upstream cov. & Lift \\
\midrule
procedure + decision & 18.7 & 7.8 & 4/6 & 1.82 \\
procedure + policy & 16.4 & 3.7 & 4/6 & 1.69 \\
lifecycle + procedure & 13.8 & 11.0 & 5/6 & 0.82 \\
decision + policy & 11.9 & 4.1 & 4/6 & 1.97 \\
lifecycle + decision & 10.0 & 8.3 & 5/6 & 0.96 \\
lifecycle + policy & 9.5 & 6.4 & 5/6 & 0.97 \\
procedure + decision + policy & 8.7 & 3.1 & 2/6 & 3.53 \\
lifecycle + retarget & 7.6 & 14.1 & 5/6 & 1.14 \\
procedure + resource & 7.5 & 4.6 & 3/6 & 1.09 \\
procedure + spec & 7.5 & 3.0 & 4/6 & 1.55 \\
\bottomrule
\end{tabular}
}
\end{table}

The most prevalent and strongly associated bundles combine behavioral families:
procedure + decision (18.7\%, lift 1.82), procedure + policy (16.4\%, lift
1.69), and decision + policy (11.9\%, lift 1.97) each hold across at least
four repositories, indicating that workflow steps, branching decisions, and
behavioral constraints are adapted together. However, their low skill breadth
(3.7\%--7.8\%) confirms that these bundles recur intensively within a small
set of skills rather than ecosystem-wide.

Lifecycle + retarget presents the contrasting pattern: modest prevalence
(7.6\%) and lift (1.14), but the highest skill breadth of any bundle (14.1\%)
spanning five repositories. Retargeting a skill to a different agent tool
naturally entails repackaging, renaming, and metadata updates, so the two
families travel together broadly rather than within a single community.

\begin{rqanswer}{RQ3}
    Skill adaptations rarely involve a single family. Procedure, decision, and
    policy form the most frequent adaptation bundles, indicating that changes
    to skill behaviour are commonly made together. In contrast, lifecycle and
    retarget exhibit the broadest coverage across skills, suggesting that these
    families are widely involved when adapting skills across repositories.
    Overall, skill adaptation is typically a multi-family activity, motivating
    the need for change-impact analysis and cross-family consistency checking
    to preserve consistency across related adaptations.
\end{rqanswer}

\subsection{RQ4. Security-Sensitive Additions}

\begin{table}[t]
  \centering
  \caption{Most common security-sensitive patterns detected in content added by downstream adaptations.}
  \vspace{-0.3cm}
  \label{tab:rq4-security-added-patterns}
  \resizebox{\columnwidth}{!}{%
  \begin{tabular}{@{}lrrrr@{}}
    \toprule
    Pattern & Instances & Branch prev.(\%) & Matches & Match share(\%) \\
    \midrule
    Access to Sensitive Files & 149 & 13.2 & 867 & 55.5 \\
    External Network Requests & 66 & 5.9 & 283 & 18.1 \\
    Global Package Installation & 40 & 3.5 & 152 & 9.7 \\
    Execution of Dangerous Commands & 28 & 2.5 & 127 & 8.1 \\
    Dynamic Code Execution & 27 & 2.4 & 58 & 3.7 \\
    System Command Invocation & 22 & 1.9 & 41 & 2.6 \\
    \bottomrule
  \end{tabular}
  }
\end{table}

Among the 1,126 skill-adaptation instances, 223 (19.8\%) match at least one
security-sensitive pattern. Of these, 209 (18.6\%) match in added lines and
14 match only in deleted lines. Since we focus on content introduced by
downstream adaptations, the following analysis covers the 209 added-line
instances.

Table~\ref{tab:rq4-security-added-patterns} shows that \textit{Access to
Sensitive Files} is the most common pattern, appearing in 13.2\% of all
adaptation instances and accounting for 55.5\% of all added-line matches.
\textit{External Network Requests} is a distant second (5.9\%), with the
remaining patterns each appearing in fewer than 4\% of instances.

Beyond which rules fire, we also examine where the matched content falls:
73.1\% of added-line matches fall in \texttt{SKILL.md} or bundled
documentation, and only 18.1\% fall in scripts. This reflects the nature
of skills, where behavioral guidance is delivered through natural-language
instructions rather than code, so security-relevant directives appear as
prose rather than executed operations. The detector rules were designed for
source code, so a textual match in a prose directive identifies a surface
to review rather than a confirmed risk or functional requirement.

\begin{rqanswer}{RQ4}
  Downstream forks do not just inherit a skill's security posture, they
  actively add to or modify it: 18.6\% of adaptation instances newly introduce content
  matching security-sensitive patterns, and access to sensitive files alone
  accounts for over half of these new matches. This new content lands mostly
  in \texttt{SKILL.md} or bundled documentation rather than scripts, beyond the reach
  of code-focused security review.
\end{rqanswer}

\subsection{RQ5. Commit-Message Coverage}

\begin{table}[t]
\centering
\caption{What and Why expression across the 277 audited commit-message sets. Each
value is the percentage of sets assigned to that category; categories are
multi-label, so percentages within a block do not sum to 100\%.}
\vspace{-0.3cm}
\label{tab:rq5-message-expression}
\resizebox{\columnwidth}{!}{%
\begin{tabular}{@{}lr@{}}
\toprule
Category & Percentage of commit-message sets (\%) \\
\midrule
\multicolumn{2}{@{}l}{\textit{How messages describe What}} \\
Summarize skill artifact change              & 92.1 \\
Illustrate skill function                    & 90.6 \\
Describe skill implementation principle      & 54.2 \\
Missing What                                 & 3.6 \\
\midrule
\multicolumn{2}{@{}l}{\textit{How messages explain Why}} \\
Describe skill objective                     & 76.2 \\
Describe skill issue                         & 45.1 \\
Illustrate skill requirement                 & 21.7 \\
Imply skill necessity                        & 15.2 \\
Missing Why                                  & 9.0 \\
\bottomrule
\end{tabular}
}
\end{table}

Two authors independently audited the agent-assigned commit-message labels,
reaching substantial agreement (Cohen's $\kappa$=0.80, Krippendorff's
$\alpha$=0.87).

Commit messages are an incomplete record of \emph{what} changed: a message
names at least one patch pattern in 84.5\% of the 277 records, yet captures
only 63.4\% of the patch labels on average, and 59.8\% of the patterns it
asserts have no match in the patch. We therefore focus on the \emph{why}:
whether the stated rationale identifies the kind of adaptation made.

The stated rationale is overwhelmingly a functional objective
(Table~\ref{tab:rq5-message-expression}): \emph{objective}, which states the
goal a change is meant to achieve, dominates at 76.2\% of records, ahead of
\emph{issue} (45.1\%), while \emph{requirement} (21.7\%) and \emph{necessity}
(15.2\%) are far less common. These objectives describe the action taken
rather than a deeper need, announcing a feature added, a platform supported, or
content translated. Developers thus record what a change accomplishes more
often than why it was needed.

\begin{figure}[t]
\centering
\includegraphics[width=\columnwidth]{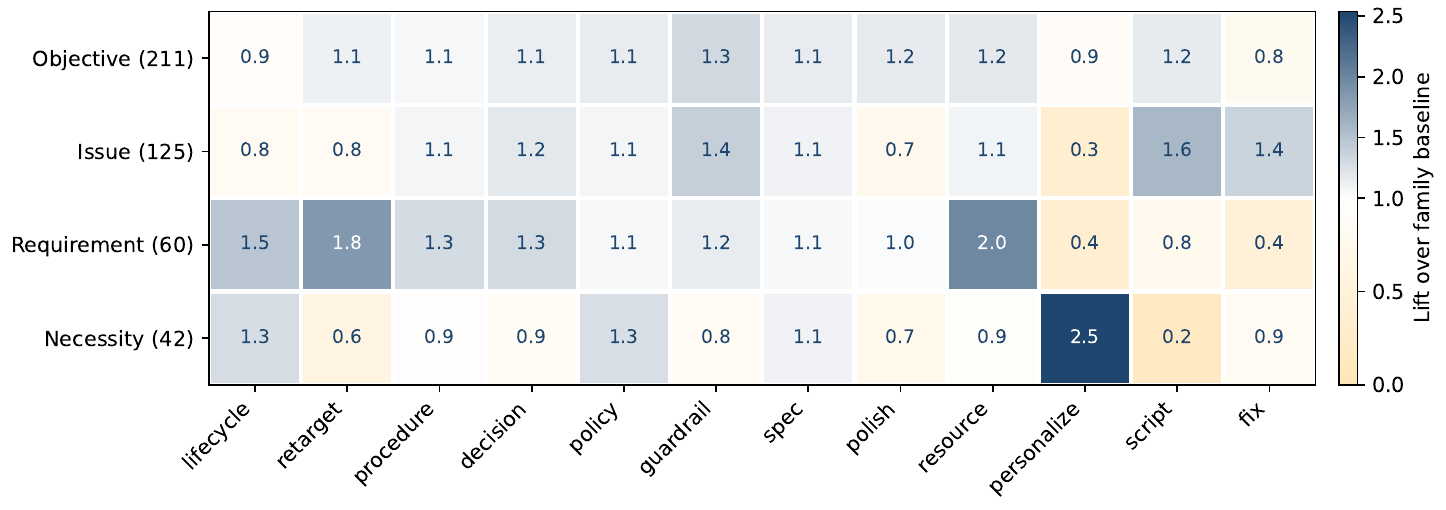}
\vspace{-0.8cm}
\caption{Relationship between commit-message Why categories and patch-derived
adaptation families. Each cell reports lift, computed as
\(P(\mathrm{family}\mid\mathrm{Why}) / P(\mathrm{family})\). A value of 1
indicates no enrichment over the family baseline, values above 1 indicate
over-representation, and values below 1 indicate under-representation.}
\label{fig:rq5-why-family-heatmap}
\end{figure}

This dominant rationale therefore does not indicate which kind of adaptation
was made: the objective category spreads almost evenly across families, with
lift staying within a narrow band (0.80--1.31) for every family in
Figure~\ref{fig:rq5-why-family-heatmap}. Discriminating signal appears only in
the rarer rationales that state a concrete need: \emph{necessity} is strongly
associated with personalize (lift 2.5), and \emph{requirement} with resource
(lift 2.0) and retarget (lift 1.8). For most commits, then, the stated
rationale identifies the goal but not the deeper customization need behind it.

\begin{rqanswer}{RQ5}
  Commit messages cannot substitute for reading the change itself: the
  dominant stated rationale, a functional objective, spreads almost evenly
  across every adaptation family and so cannot indicate which kind of
  adaptation was made. Only the rarer necessity and requirement rationales
  carry any discriminating signal. A message-based proxy would therefore
  mislead a researcher trying to infer the customization a skill underwent,
  which is why our taxonomy relies on the diff itself, not stated intent.
\end{rqanswer}

\section{Discussion and Implications}\label{sec:discussion}
This section discusses the implications of our findings to different stakeholders, as well as a simple RADAR checklist we provide to help downstream skill developers.



\subsection{Implications for Researchers}

\paragraph{Use adaptation patterns as variables for studying skill improvement.}

Downstream skill evolution consists of recurring adaptation types.
Existing self-evolving skill frameworks optimize or evaluate updated skills
as complete artifacts~\cite{linMUSEAutoskillSelfEvolvingAgents2026,
yangSkillOptExecutiveStrategy2026}, but our taxonomy makes each adaptation
type analyzable. Researchers can annotate agent trajectories, fork histories,
or self-evolution logs by adaptation type and examine which changes improve
downstream task behavior, shifting the question from whether a skill improved
to which kinds of adaptations produced the improvement.

\paragraph{Infer customization needs from change content, not stated intent.}

A natural way to study why skills are customized is to read the intent
developers state in commit messages, but RQ5 shows this rationale is too
coarse to recover the customization need behind a change. Researchers
studying skill evolution should instead infer needs from the content of
the modification, as our taxonomy does, and treat commit messages and
other narrative metadata as a weak auxiliary signal rather than ground
truth for intent~\cite{tianWhatMakesGood2022}.

\subsection{Implications for Harness Engineers}

\paragraph{Provide native support for recurring customization needs.}

The recurring needs in RQ1, namely making a skill discoverable, controlling its
behavior, and retargeting it, are met today by manually rewriting prose, so
skill engineering tools could support them as first-class operations rather than
free-text edits. RQ3's two coupling patterns point to concrete leverage: the
coordinated behavioral edits to workflow steps, decision rules, and policy
constraints suggest tooling that exposes configurable decision logic and flags
when one edit should propagate to the others, while the broad structural
coupling between retargeting and repackaging suggests a retarget-aware tool
that automates the repackaging, renaming, and \texttt{description} rewrite
that porting otherwise forces developers to perform by hand.

\paragraph{Make the automation and security tradeoff explicit.}

Expanding skill capability often introduces security-relevant content as a
byproduct. Local context access, network use, command execution, and dynamic
execution make skills more useful while expanding the permissions they
require~\cite{liuAgentSkillsWild2026,liNoAttackRequired2026}. RQ4's
sensitive-file-access finding (Section~\ref{sec:results}) is a case in point.
Because these detections are based on textual pattern matching rather than
behavioral analysis, they identify review surfaces rather than confirmed risks. Skill
platforms should therefore surface security-sensitive changes during
adaptation, enforce least-privilege execution, and apply automated checks
over scripts, files, and referenced resources.

\subsection{Implications for Downstream Skill Developers and the RADAR Checklist}

The findings from RQ1 to RQ5 suggest that downstream skill adaptation
should be managed as a coordinated modification activity rather than
isolated prompt editing. These observations motivate a lightweight
developer-facing checklist, \textbf{RADAR}: \textbf{R}oute the skill by
checking activation metadata, \textbf{A}dapt behavioral logic across
procedures, decisions, and policies, \textbf{D}eclare dependencies among
resources and scripts, \textbf{A}udit security-sensitive instructions and
execution surfaces, and \textbf{R}ecord changes for future maintenance, an
implication-driven review lens rather than a validated technique.

\section{Threats to Validity}\label{sec:threats}

We discuss the principal threats to the validity of our findings, organized
according to the standard categories of construct, internal, external, and
conclusion validity, and describe the measures taken to mitigate them.

\paragraph{Construct Validity.}

Our measures may not fully capture the breadth of skill adaptation.
Public forks provide only observable evidence of modifications that
contributors choose to publish; adaptations made in private or
abandoned forks are absent from the corpus.
Counting each adaptation pattern once per branch instance may also
underrepresent repeated edits within the same variant; we mitigate
this by reporting presence rather than frequency.
The security analysis identifies potential risks through pattern
matching and should not be interpreted as confirming vulnerabilities;
our findings characterize modification surfaces and review burden
rather than exploitable flaws.

\paragraph{Internal Validity.}

The assigned labels may be influenced by the labeling agent.
To mitigate this, each label required supporting evidence and a
confidence assessment, supplemented by human review and consensus
reconciliation. A related concern is that auditors reviewed
agent-assigned labels rather than labeling each patch from scratch,
which could anchor them to the agent output and inflate
consensus-vs-agent agreement. The reconciliation record indicates this
did not dominate, as auditors independently added labels the agent had
omitted, a pattern inconsistent with simply accepting the agent output;
we therefore treat inter-auditor agreement as the primary reliability
estimate.

\paragraph{External Validity.}

Our findings may not generalize beyond the studied repositories.
The corpus derives from six popular repositories selected from a
single GitHub snapshot, and one repository accounts for a
disproportionately large share of branch instances. Because GitHub
search results are time-varying, the exact pre-screening candidate
counts cannot be reconstructed from the snapshot we collected. To
mitigate these limitations, we release the search scripts, filtering
thresholds, and downstream mining outputs to make the selection
procedure inspectable, and report repository-adjusted prevalence and
upstream coverage metrics to reduce the influence of any single large
repository.

\paragraph{Conclusion Validity.}

The conclusions may depend on the model used for labeling. The labeling
harness did not expose a stable model snapshot identifier or temperature
control; the only configurable generation setting was reasoning effort, fixed
to medium throughout. All reported analyses use the human-reconciled consensus
labels for audited records to mitigate the impact of labeling errors, and we
release the labeling skill, audit records, and analysis scripts to support
independent verification and replication.

\section{Conclusion}\label{sec:conclusion}

We present the first empirical study of downstream skill adaptation, mining
1,126 skill-adaptation instances from six widely adopted public skill
repositories and constructing a validated taxonomy of 46 adaptation
patterns grouped into 13 families. Our results show that a published skill
is rarely adopted as is: developers rewrite it to restore discoverability,
control its behavior, and retarget it to a different agent tool or
language, work concentrated in \texttt{SKILL.md} and coupled across families in
ways that call for change-impact and consistency checking. A substantial
share of this rewriting introduces security-sensitive content within the
same instruction text that governs behavior, while the commit messages
that record these changes capture what was done far more reliably than
why, limiting their use as a proxy for the underlying customization need.
These findings carry implications for skill formats, management tooling,
and the security of reused skills.

\section{Data Availability}
\label{sec:data-availability}

The artifact is available online at \textcolor{blue}{\url{https://anonymous.4open.science/r/skill-pattern-mine-1BDE/README.md}}.
It includes the repository-mining materials, agent configuration, taxonomy
definitions, taxonomy-building and labeling prompts, labeled corpus, audit
materials, analysis scripts, and security-pattern definitions needed to inspect
and reproduce the analyses reported in this paper.

\clearpage
\bibliographystyle{IEEEtran}
\bibliography{skills}


\end{document}